\documentclass{osa-article}

\journal{osac}
\articletype{Research Article}
\usepackage{mathtools} %words below arrow
\usepackage{subcaption}
\usepackage{graphicx}
\usepackage{ulem}

\usepackage{color}
\def\revb{\color{black}}
\def\OSAC{\color{black}}

\def\colorb{\color{black}}
\def\cb{\color{black}}
\def\colorbb{\color{black}}

\def\cred{\color{black}}

\renewcommand{\sout}[1]{\unskip}
\usepackage{soul}

\def\G{\boldsymbol{G}}

\def\r{\boldsymbol{r}}

\def\v{\boldsymbol{v}}
\def\G{\mathsf{G}}
\def\U{\mathsf{U}}

\def\rot{\boldsymbol{\nabla}}

\def\ep{\varepsilon}
\def\om{\omega}
\def\R{\mathbb{R}}

\begin{document}

\title{Completeness and divergence-free behavior of the quasi-normal modes using causality principle}

\author{M. Ismail Abdelrahman\authormark{1,2} 
and B. Gralak\authormark{1,3} }

\address{\authormark{1}Aix Marseille Univ, CNRS, Centrale Marseille, Institut Fresnel, Marseille, France}
%\authormark{*}m.abdelrahman@fresnel.fr\\
\email{\authormark{2}m.abdelrahman@fresnel.fr \\
\authormark{3}boris.gralak@fresnel.fr} %% email address is required

\homepage{https://doi.org/10.1364/OSAC.1.000340} %% author's URL, if desired


\begin{abstract}
A fundamental feature of the quasi-normal modes {\cred{(}}QNMs{\cred{)}}, which describe light interaction with  open (leaky) systems like nanoparticles, lies in the question of the completeness of the QNMs representation and in the divergence of their field profile due to their leaky behavior and complex eigenfrequency. In this article, the QNMs expansion is obtained by taking into consideration the frequency dispersion and the causality principle. The derivation based on the complex analysis ensures the completeness of the QNMs expansion and prevents from any divergence of the field profile. The general derivation is tested in the case of a one-dimensional open resonator made of a homogeneous {\colorb{absorptive
}} medium with frequency dispersion given by the Lorentz model. For a harmonic excitation, the result of the QNMs expansion perfectly 
matches the exact formula for the field distribution outside as well as inside the resonator.
\end{abstract}

%%%%%%%%%%%%%%%%%%%%%%%%%%  body  %%%%%%%%%%%%%%%%%%%%%%%%%%
\section{Introduction}

Light interaction with nanoparticles enables a wide range of unprecedented applications such as high-resolution spectroscopy, photothermal cancer therapy, optical tweezers, and light steering using  metasurfaces \cite{nie1997probing,huang2006cancer,grigorenko2008nanometric, ni2012broadband}. The nanoparticles can be viewed as open resonators that are able to confine light down to subwavelength dimensions. Since the energy can escape towards the
surroundings, \sout{of these resonators,} such open systems are nonconservative
and their behavior can be described by quasi-normal modes (QNMs) characterized by complex eigenvalues \cite{leung1994completeness1,leung1994completeness2, ching1998quasinormal}. 

The QNMs expansion provides  direct physical description and interpretation of the light-particle interaction through  the investigation of the resonant modes of the system. In addition, the QNMs expansion may lead to reduced computational efforts
since the resonant modes of the system have to be calculated only once and then used to solve the excitation problem by evaluating the contribution of the \sout{few} resonances {\colorb{modes}} near the excitation frequency. The QNMs approach has shown its relevance in many situations such as the calculation of the effective mode volume for leaky cavities \cite{kristensen2012generalized} and of the Purcell factor \cite{sauvan2013theory}, the modeling of light behavior in plasmonic nanoparticles \cite{ ge2014quasinormal, ge2014design}, the design of  angle-independent spectral filters \cite{ vial2014resonant}, the scattering matrix calculations \cite{ alpeggiani2017quasinormal},  the local density of states  calculations \cite{ de2015semianalytical}, the modeling of the perturbations of black holes \cite{nollert1992quasinormal}, and the list goes on. Moreover, it has been recently demonstrated 
  that the peaks of the extinction spectrum of a scattering particle are associated with its QNMs \cite{ powell2017interference}, which cannot be elucidated by the multipolar expansion.

The QNMs expansion suffers however from a longstanding limitation: the complex nature of the eigenvalues $z_q$ of  open systems results in the divergence of the field outside the resonator, i.e. $e^{i z_q |\r|} \rightarrow \infty$ when 
$|\r| \rightarrow \infty$. This divergence complicates the normalization of the QNMs since the energy of the modes is unbounded and, more importantly, the QNMs expansion appears to fail in describing the radiation outside the open resonator. Several solutions have been proposed to explicitly mitigate the normalization difficulty for leaky resonators 
\cite{kristensen2012generalized, sauvan2013theory,bai2013efficient}. \sout{{\cb{Other approaches based on the Dyson equation formalism \cite{ge2014quasinormal} lead to the definition of regularized QNMs 
from the knowledge, inside the resonator, of standard QNM which are then propagated in the background using the Green's function of the background \cite{PhysRevB.97.115302}.}} \sout{However, to the best of our knowledge} The validity of the QNMs expansion outside the resonator remains {\cb{however}} 
an open problem that questions {\cb{the completeness of the QNMs, i.e.}} 
the ability of the QNMs to form a complete set of modes to represent the field in the whole space{\cb{, especially in the case of absorptive and dispersive 
systems}}. A recent review could be consulted for an exhaustive discussion on the recent advances in the topic of QNMs expansion \cite{lalanne2017light}. {\colorb{ \sout{@Discussion: I think we should also briefly mention other methods than QNM to describe the diffracted fields of the open systems} \cb{I think it is not necessary, we can focus on QNMs}}}.
In Ref. \cite{vial2014quasimodal}, it was suggested that the exponential divergence outside the resonator is attributed to a noncausal effect (excitation at past times). {\colorb{Herein}} }
%%%%%%%%
{\colorbb{Other approaches based on the Dyson equation formalism \cite{ge2014quasinormal} lead to the definition of regularized QNMs from the knowledge of standard QNM inside the resonator, which are then propagated in the background using the Green's function of the background  \cite{PhysRevB.97.115302}. The theoretical question of the validity of the QNMs expansion outside the resonator still remains   an open problem that doubts the completeness of the QNMs, i.e., the ability of the QNMs to form a complete set of modes to represent the field in the whole space, especially in the case of absorptive and dispersive systems. In Ref. \cite{vial2014quasimodal}, it was suggested that the exponential divergence outside the resonator is attributed to a noncausal {\colorbb{flaw in the definition of the QNMs}}. A recent review could be consulted for an exhaustive discussion on the recent advances in the topic of QNMs expansion \cite{lalanne2017light}.}}

In this article, the QNMs expansion is derived taking into consideration the frequency dispersion and the 
causality principle. This approach has recently demonstrated the validity of the modal expansion to represent the transient fields produced by a one-dimensional resonator \cite{abdelrahman2018modal}. \sout{In the present case} {\colorb{Herein}}, it is shown that an additional causality-related phase factor cancels the exponential {\cred{divergence}} of the {\cred{``conventional''}} QNMs {\colorb{ \sout{hence  enables the construction of the true {\cred{(causal)}} QNMs} of  open systems}}. Arguments are proposed to support the validity of the expansion on the \sout{causal-} QNMs of the Green's function in  the whole space. A simple example is presented for a one-dimensional resonator made of a homogeneous {\cb{absorptive}} dispersive medium, where the Green's function of the resonator is expanded using  QNMs and then the obtained results are compared to the exact formulation. 

\section{Causality of the Green's function}

The electromagnetic Green's function {\revb{$\G(\r,\r';z)$}} of a system of permittivity 
$\ep(\r,z)$ is defined as the solution {\revb{of}} the Helmholtz equation for a {\revb{point Dirac 
current source $\delta(\r-\r')$ located at $\r'$:}}
\begin{equation} \label{eq:Helm}
%\begin{array}{r}
{\cb{ \rot \times \rot \times \G(\r,\r';z) - z^2 \mu_0 \, \ep(\r,z) \, \G(\r,\r';z) 
 = i z \mu_0  {\revb{\, \U \, \delta(\r-\r')}} \, .}}
%\end{array}
\end{equation}
{\revb{Here the following notations have been adopted: $\r$ (and $\r'$) is the position 
vector of space, $\rot \times$ is the curl operator, $z$ is the complex frequency, 
$\mu_0$ is the vacuum permeability, and $\U$ is the unit dyadic tensor.}} {\colorb{The reference system {\cb{is defined as}} the infinite media surrounding the resonator {\cb{and}} is characterized by the permittivity $\ep_{\text{ref}}(\r,z)$ and the Green's function $\G_{\text{ref}}(\r,\r';z)$.}} 
The difference $\Delta \ep(\r,z) = \ep(\r,z) - \ep_{\text{ref}}(\r,z)$ 
corresponding to the resonator must be restricted to a bounded domain. {\revb{Let $\G_d (\r,\r';z)$ be the ``diffracted'' (or ``difference'') 
Green's function, corresponding 
to the field diffracted by the resonator, defined as the difference}}

\begin{equation}
%\begin{array}{ll}
\G_d (\r,\r';z)  = \G(\r,\r';z) - {\revb{\G_{\text{ref}}(\r,\r';z)}} \, . %\\[1mm]
%& = - i z \mu_0 \left\{ \Big[ \He_1(z) \Big]^{-1} - \Big[ \Hv(z) \Big]^{-1} \right\} \deltabold(\r',z) \\[2mm]
%& = \,\, i z^3 \mu_0 \Big[ \He_1(z) \Big]^{-1} \,\Delta\ep\, \Big[ \Hv(z) \Big]^{-1} \deltabold(\r',z)
%\label{eqEd}
%\end{array}
\end{equation}
%where $ \Delta\ep= \Big(\ep_1(\r,z)-\ep_0(\r,z)\Big)$. \\
For dispersive media, {\revb{the permittivities $\ep(\r,z)$ and $\ep_{\text{ref}}(\r,z)$
tend to the vacuum one $\ep_0$ when the frequency $z \rightarrow \infty$,}} which in 
turn suggests 
{\revb{when $z \rightarrow \infty$  that }}
\begin{equation} \label{eq:Conv}
{\revb{\G(\r,\r';z) \rightarrow \G_{\text{ref}}(\r,\r';z) 
\rightarrow \G_0 (\r-\r';z) \: \Longrightarrow \: \G_d (\r,\r';z) \rightarrow 0 \, ,}}
\end{equation}
{\revb{where $\G_0 (\r-\r';z)$ is the free Green's function in vacuum.}}

The {\revb{Fourier transform with respect to the time variable $t$ of the diffracted 
Green's function can be defined as}}
\begin{equation}
\widehat{\G}_d(\r,\r';t) = \int_{\Gamma_\eta} dz \,e^{-izt} \G_d(\r,\r'; z) \, ,
\label{eqGt}
\end{equation}
{\revb{where $\Gamma_\eta$ is the line parallel to the real axis of complex frequencies 
$z = \omega + i \eta$, with real part $\omega$ describing $\R$ and imaginary part 
set to the positive number $\eta$. It is stressed that the integral (\ref{eqGt}) 
is well-defined for all $t$, which results from the following property 
 of the permittivity \cite{gralak2017analytic}: 
$\int_\R d \omega \big| \ep(\om+i \eta ) - \ep_0 \big| < \infty$. }} The resulting time-dependent function (\ref{eqGt}) is the Green's function generated
by the spatio-temporal Dirac source $\delta(\r-\r') \delta(t)$.

{\revb{The evaluation of the integral expression (\ref{eqGt}) is based on the remark 
(\ref{eq:Conv}) stating that the Green's functions $\G(\r,\r';z)$ and 
$\G_{\text{ref}}(\r,\r';z)$ tend to 
the free one $\G_0(\r,\r';z)$ when $z \rightarrow \infty$, so that \sout{their}  {\cred{ the }}  exponential 
behavior  {\cred{ of the Green's functions $\G(\r,\r';z)$ and 
$\G_{\text{ref}}(\r,\r';z)$  }} at high frequencies is governed by {\cb{$e^{i z |\r - \r'| / c} $,}}  
where $c = 1 / \sqrt{\ep_0 \mu_0}$ is the light velocity in vacuum. Hence it appears 
reasonable to make this first assumption $(\textrm{I})$:
\begin{equation} 
\G_d (\r,\r';z) \times \, e^{-i z \tau}  \, e^{-i z |\r - \r'| / c} 
\underset{z \rightarrow \infty}{\longrightarrow} 0 \, . 
\label{normGd}
\end{equation}
When the complex frequency $z$ has a positive imaginary part, this assumption 
is always true for $\tau = 0$, while when $z$ has negative imaginary part, 
it could be necessary to consider {\cb{the arbitrary small time 
 $\tau > 0$,}} to ensure the vanishing behavior at the limit $z \rightarrow \infty$. A second assumption $(\textrm{II})$ is based on the bound nature of the resonator and of the difference of permittivities 
$\Delta \ep(\r,z) = \ep(\r,z) - \ep_{\text{ref}}(\r,z)$. Under this condition, it 
can be assumed that the diffracted Green's function $\G_d(\r,\r'; z)${\OSAC{, which corresponds to the scattering operator \cite{PRB14}, }} has only 
a discrete set of resonances $\{ z_q \}$ located in the lower half plane of complex frequencies {\cb{(arguments based on the analytic Fredholm theorem \cite{reed1979methods} {\OSAC{or a boundary intergral expression of the scattering operator \cite{PRB14}}} can be used to support this assumption)}}. 

The time-dependent Green's function (\ref{eqGt}) is evaluated considering 
the two following cases. For $t <  |\r - \r'|/c$, the line $\Gamma_\eta$ is 
deformed in the upper half plane of complex frequencies, i.e. with positive imaginary 
part, leading to $\widehat{\G}_d(\r,\r', t) =0$, 
since all the Green's functions are 
analytic in this domain. For $t > \tau + |\r - \r'|/c$, the line $\Gamma_\eta$ is 
deformed in the lower half plane and the Green's function is given by the 
contributions of the {\colorb{set of}} poles $\{z_q\}$:
\begin{equation} \label{GdResidue}
\widehat{\G}_d(\r,\r', t) = - 2 i \pi \displaystyle\sum_{\{ \text{poles } z_q\} } 
 e^{-i z_q t}\, \G_d^q(\r,\r', z_q) \, ,
\end{equation}
where $\G_d^q(\r,\r', z_q)$ is the residue of the function 
$\G_d(\r,\r', z)$ at its pole $z_q$.}}
The application of the residue theorem is justified {\colorb{by the assumption $(\textrm{I})$}} that ensures the vanishing value of the integral (\ref{eqGt}) 
on an infinite semi-circle in {\revb{the}} lower half plane.

{\revb{Finally, the harmonic Green's function is retrieved by applying the 
inverse Laplace transform to the expression (\ref{GdResidue}). 
For an arbitrary small $\tau$  and for a frequency $z$ with positive imaginary part, 
the following function is defined:
\begin{equation} 
\label{Gexpansion}
%\begin{array}{ll}
\widetilde{\G}_{\tau,d}(\r,\r'; z) = \dfrac{1}{2\pi} 
\displaystyle\int^\infty_{\tau + |\r-\r'|/c} dt \: e^{i z t} \, 
\widehat{\G}_d(\r,\r', t) \, .%\\[4mm]
%& = \displaystyle\sum_{\{ \text{poles } z_q \} } 
%\dfrac{\Res_d(\r,\r',z_q)}{z - z_q}\: 
%e^{i (z - z_q) \,(\tau + |\r-\r'| / c)}  \\[4mm]
%\end{array}
\end{equation}
{\cb{Notice that the arbitrary small time $\tau$, introduced to ensure the 
vanishing limit (\ref{normGd}), leads to avoid  the time $\tau$ at 
the {\colorbb{start}} of the time-dependent Green's function in the Laplace transform 
(\ref{Gexpansion}).}}
It can be shown that the {\cb{resulting}} 
function $\widetilde{\G}_{\tau,d}(\r,\r'; z)$ tends to 
the required Green's function ${\G}_{d}(\r,\r'; z)$ when $\tau$ tends to zero
since the time-dependent function $\widehat{\G}_d(\r,\r';t)$ is {\cb{{\cred{bounded}} with respect to time $t$\sout{for the case of a disperive medium, i.e. $\widehat{|\G}_d(\r,\r';t)| < \infty$ for all $t$}}}. Hence the diffracted Green's function can be expressed as an expansion 
of the complex resonances (or QNMs):
\begin{equation} 
\label{G-QNM}
\G_{d}(\r,\r'; z) = \displaystyle\lim_{\tau \rightarrow 0} 
\displaystyle\sum_{\{ \text{poles } z_q \} } 
\dfrac{\G_d^q(\r,\r', z_q)}{z - z_q}\: 
e^{i (z - z_q) \,(\tau + |\r-\r'| / c)}  \, .
\end{equation}
{\cred{This QNMs expansion is  derived using complex analysis while taking into consideration the causality principle that appears in the lower limit of the integral in Eq. (\ref{Gexpansion}), which ensures that this QNMs expansion cannot lead to an exponential increasing behavior.}} This can be checked remarking that each residue $\G_d^q(\r,\r', z_q)$ 
is damped by the factor $e^{ i (z - z_q )|\r-\r'| / c}$, which allows the representation of the diffracted field  in the whole space, inside and outside the resonator. {\colorb{Furthermore,}} the  {\OSAC{set of QNMs of the scattering operator forms a complete set for the diffracted field since the Green's function has been expanded using solely these modes and without the continuum associated to the infinite reference system.}} However, as it will be observed afterwards, the convergence 
of the QNMs expansion above may appear  slow for $\tau \rightarrow 0$. 
This may require a special attention to  use  a reasonable number 
of QNMs in practice. 
}}
%This additional causality-related term cancels the exponential growing term in the residue expression  and hence ensures no exponential growing.

\section{Test Case: 1-D dispersive resonator}
%------------------------
%  Fig 1
%------------------------
\begin{figure}[h!]
\centering
\includegraphics[ width=0.9\linewidth, keepaspectratio]{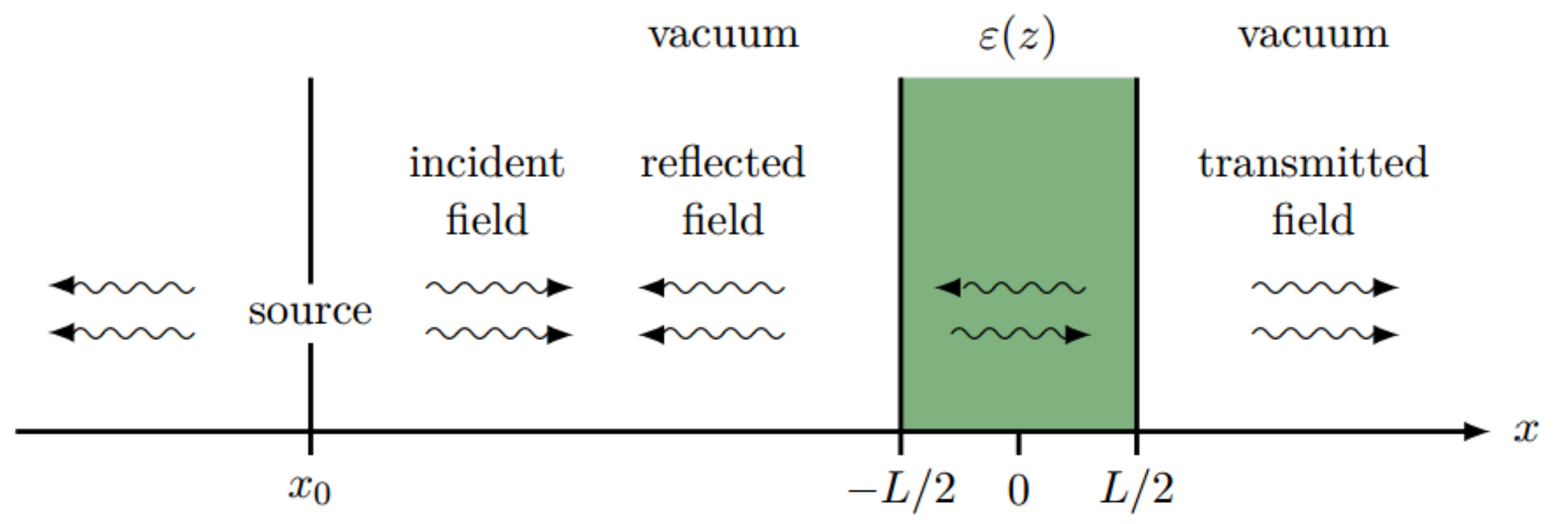}
\caption{A one-dimensional resonator made of a homogeneous dispersive material is excited by a harmonic source located at $x_0$ \sout{{\colorb{(in a normal incidence)}}}. This interaction results in diffracted fields both in the forward (transmission) and backward (reflection) directions.} \label{Fig1}
\end{figure}
%-----------------------

{\colorb{To test the validity of the derived QNMs expansion}} a simple one-dimensional example is shown in Fig. \ref{Fig1} for a  homogeneous 
 medium of thickness $L$ centered at {\colorb{the origin}} $x=0$, with a relative 
{\cred{permittivity}} {\colorb{  given by the Lorentz model 
%\begin{equation}
$\ep(z)= 1- \Omega^2/(z^2 %{\abs{+i\gamma z}}
-\omega_0^2+iz\gamma)$,
%\end{equation}
where $\omega_0$ is the medium resonance frequency, $\gamma$ is the absorption parameter, and  $\Omega$ is related to the electron density. This medium is surrounded by vacuum and hence  acts as a {\cb{Fabry-P{\'e}rot 
resonator\sout{ due to the multiple reflection at the interfaces causes by the discrepancy of the permittivity}}}.}}

The resonator is excited by a harmonic source of frequency $z$ located at $x_0$ which 
generates the incident electric field $E_{s}(x,z)= e^{i z |x-x_0|/c} 
{\cb{e^{iz(x_0+L/2)/c}}}$, normalized to unity at the input interface 
{\cb{$x=-L/2$}} of the resonator. The total 
electric field {\colorb{in the whole space {\cb{is given by the exact 
formulation \sout{can be evaluated using the exact Airy's formulation}}} \cite{siegman1986lasers}}:}
\begin{equation} \label{Gx}
\hspace*{-0.5mm}
\begin{array}{rlr}
 E_R(x,z) = \hspace*{-2mm} & {\cb{E_{s}(x,z) }} + R(z)\, e^{-i z (x+L/2)/c } \quad  \quad & x \leq {-L}/{2} \, , \\[0mm]
\hspace*{-2mm}E_{in}(x,z) = \hspace*{-2mm}& A(z) \, e^{iz\sqrt{\ep(z)}x/c} + 
 B(z) \, e^{-iz\sqrt{\ep(z)}x/c} \quad  \quad & |x| \leq L/2 \, , \\[0mm]
E_T(x,z) = \hspace*{-2mm}& T(z)\, e^{iz (x-L/2)/c} \quad  \quad &  x \geq L/2 \, , \\[2mm]
\end{array}
\end{equation}
where $T(z)$ and $R(z)$ are the transmission and the reflection coefficients of the dispersive resonator, and $A(z)$ and $B(z)$ can be uniquely determined by matching the fields at the two  resonator interfaces. The Green's function of the resonator can hence be analytically obtained for the  case presented in Fig. \ref{Fig1}. To begin with, the transmission function is given as  \cite{siegman1986lasers}
\begin{equation} \label{eqSlabTF}
T(z) = \frac{[ 1 - r_0^2(z) ] \, e^{iz \sqrt{\ep(z)} L/c }}{1-r_0^2(z)  \, 
e^{2 iz\sqrt{\ep(z)}L/c}}, \quad
r_0(z)= \frac{1-\sqrt{\ep(z)}}{1+\sqrt{\ep(z)}} \, .
\end{equation}
{\cb{It has been proven  }}that $T(z)$ is an even 
function of the square root of the permittivity \cite{abdelrahman2018modal}, hence it contains no {\cred{branch}}-cut. 
{\cb{Then, in}} order to use the QNMs expansion (\ref{G-QNM}), the conditions 
(\ref{eq:Conv}) and (\ref{normGd}) have to be 
satisfied. {\cb{For the transmitted part of the
field, the general derivation is applied to the function}}
\begin{equation} \label{eq:Tnew}
G_T(x,z) = \frac{T(z)}{z} \: e^{iz(x-L/2)/c} \quad \quad x \geq L/2 \, .
\end{equation}
{\cb{\sout{This definition is consistent with the Helmholtz equation  (\ref{eq:Helm}) where 
the factor $z$ has been removed from the right side. That {\colorb{ensures the}} 
convergence of the QNMs expansion{\colorb{, ( at the expense of ** we need to discuss that)}}.}}} It can be checked that the expression (\ref{eq:Tnew}) 
satisfies (\ref{eq:Conv}) and (\ref{normGd}) {\cb{and thus}} the validity of  (\ref{G-QNM}) is guaranteed.  The set of 
poles $\{0,\, z_q: T^{-1}(z=z_q)= 0 \}$ and the corresponding residues $T^q(z_q)$ have been 
estimated and computed in Ref. \cite{abdelrahman2018modal}. The transmitted field expansion is then
\begin{equation} \label{eq:ETexpbefore}
\begin{array}{rl}
E_T(x,z) = &  \hspace*{-2mm} z \displaystyle\sum_{z_q  }  \frac{T^q(z_q) \, 
e^{iz_q (x-L/2)/c  }}{z_q \, (z-z_q)}  \, 
e^{i (z-z_q) (x+L/2)/c }\\[2mm] + & \hspace*{-2mm}T_0 \, e^{i z (x+L/2)/c } \: , 
\end{array}
\end{equation}
where $T_0=T(z=0)=1$. Hence it is found that the series above equals the difference 
$E_T(x,z) - e^{i z (x+L/2)/c }$ which is precisely the diffracted field in the half-space 
$x>L/2$. It can be noticed that the additional causality-related factor cancels the exponential growing term in the residue expression which ensures the absence of 
exponential growing: the terms that describe the transmitted field outside the resonator (of $x$-dependency) appear as a function oscillating {\cred{at}} the excitation frequency $z$ only, 
as in \cite{PhysRevB.97.115302}:
\begin{equation} \label{eq:ETexp}
%\begin{array}{lcl}
E_T(x,z) -  e^{i z (x+L/2)/c } =  z \sum_{z_q} \frac{T^q(z_q) \,
e^{-iz_q L/c  }}{z_q \, (z-z_q)}  \,  e^{i z (x+L/2)/c }. 
%\end{array}
%\end{array}
\end{equation}
Similarly, the reflection coefficient is given by \cite{siegman1986lasers}
\begin{equation} \label{eq:Rslab}
R(z)=  r_0(z)\,  \frac{  1-  e^{2 i z \sqrt{\ep(z)}\, L/c} }{1-r_0^2(z) \, e^{2 i z \sqrt{\ep(z)} \, L/c }} \, .
\end{equation}
The reflection coefficient $R(z)$ is also an even function of $\sqrt{\ep(z)}$ and 
hence its spectrum is restricted to a discrete set of resonances. {\cb{For the reflected part of the field, the following function is considered}}
\begin{equation}
G_R(x,z)= \frac{R(z)}{z}\, \, e^{-iz\,(x+L/2)/c}\, e^{-iz {\tau}}\quad\quad x \leq -L/2 \, .
\end{equation}
The infinitesimal {\cb{time}} $\tau >0$ is introduced to ensure 
{\cb{that}} the conditions (\ref{eq:Conv}) 
and (\ref{normGd}) {\colorb{are satisfied}}. Therefore, the QNMs expansion (\ref{G-QNM}) is valid  and  
the reflected field expansion is
\begin{equation} \label{eq:Rexp}
%\begin{array}{lcl}
E_R(x,z) = {\cb{E_s(x,z)}} + z \sum_{z_q} \frac{R^q(z_q)}{z_q \, (z-z_q)}  e^{-i z (x+L/2)/c } \, 
e^{i(z-z_q)\tau } %\\[2mm]
%&+& 
\, ,
%\end{array}
\end{equation}
where the poles ${z_q}$ of the function $R(z)$ are the same as $T(z)$, since they share the same 
denominator, and $R^q(z_q)$ are the residues of the function $R(z)$ at its poles. Notice that 
the pole at $z = 0$  has no contribution since $R(0) = 0$. The expression (\ref{eq:Rexp}) shows 
no exponential growth for the reflected fields.

\begin{figure}[h!]
\centering
\includegraphics[ width=0.9\linewidth, keepaspectratio]{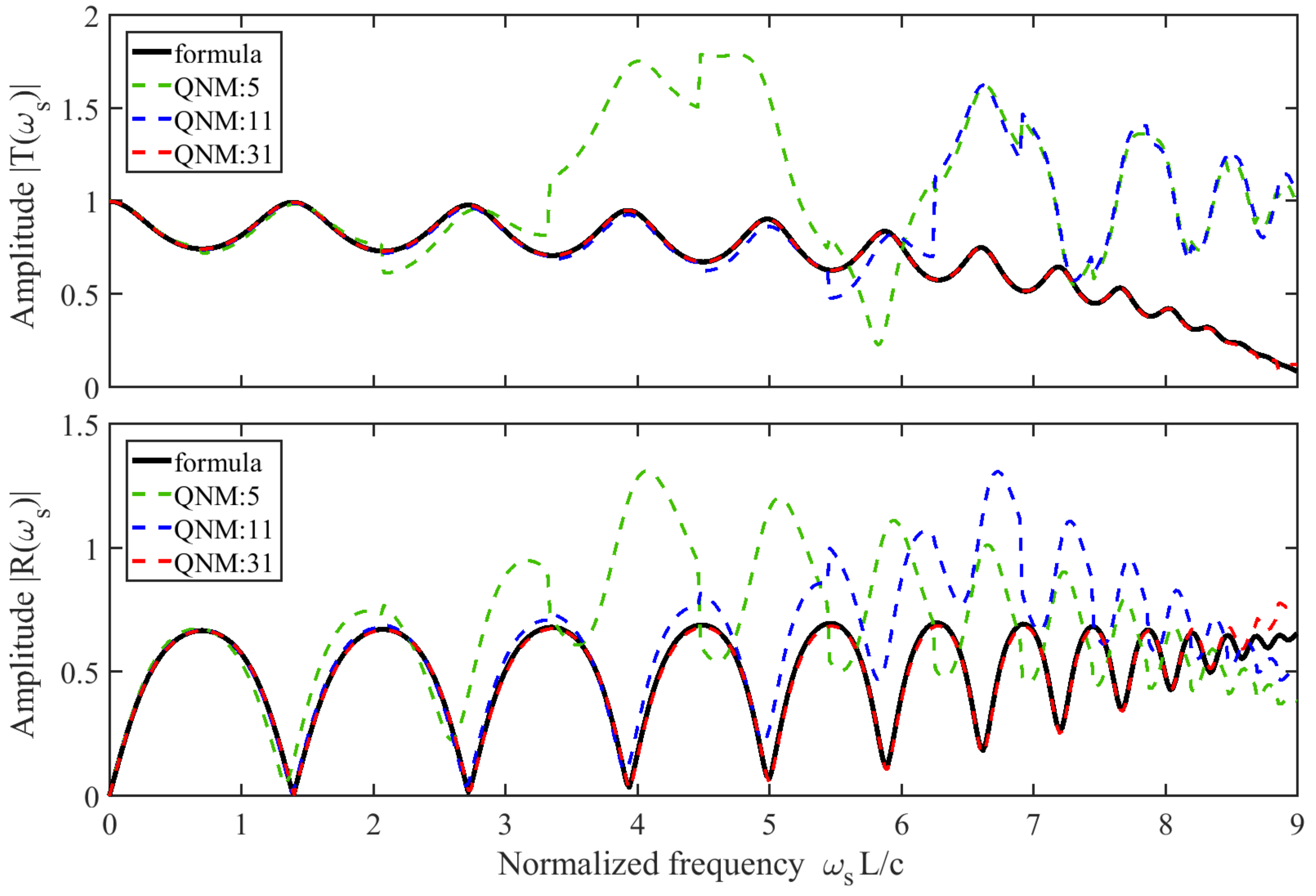}
\caption{A comparison between the results of the exact formulas {\colorb{(solid lines)}} of  $T(z)$ and   $R(z)$ and their corresponding QNMs expansion (dashed lines) for below-resonance excitation $\omega_s < \omega_0$. The results are shown for different numbers of summation terms  of the expansion.} \label{FigTR}
\end{figure}
%-----------------------

{\colorb{Figure \ref{FigTR} shows the transmission function $T(z)$ and the reflection function $R(z)$ both using the exact 
{\cred{ {\cb{expression}} (\ref{Gx}) }}and the QNMs expansion  (\ref{eq:ETexpbefore}) at $x=L/2$ for $T(z)$, and (\ref{eq:Rexp}) at $x=-L/2$ for $R(z)$ while subtracting the incident field term. The given test case is for a Lorentz medium of  parameters $\omega_0 L / c =10$, {\colorb{$\gamma L/c = 0.2$}}, and $\Omega L / c=20${\cred{, and for below-resonance
excitation with unity amplitude $|E_{s}|=1$}}. The parameter {\OSAC{$\tau c / L$ }} is set to 0.1 for the presented simulation.

{\colorb{ The convergence analysis is also presented while the results of the exact formulation and the QNMs expansions calculations}} show an excellent agreement, if enough summation terms are included. {\colorb{As mention earlier, this parameter $\tau$ ensure the condition (\ref{normGd}) at the expense of introducing a {\cb{small}} error. It is possible to decrease the value of $\tau$, however, the number of modes needed to ensure the convergence is increased accordingly.  }}

\begin{figure}[h!]
\centering
\includegraphics[ width=0.9\linewidth, keepaspectratio]{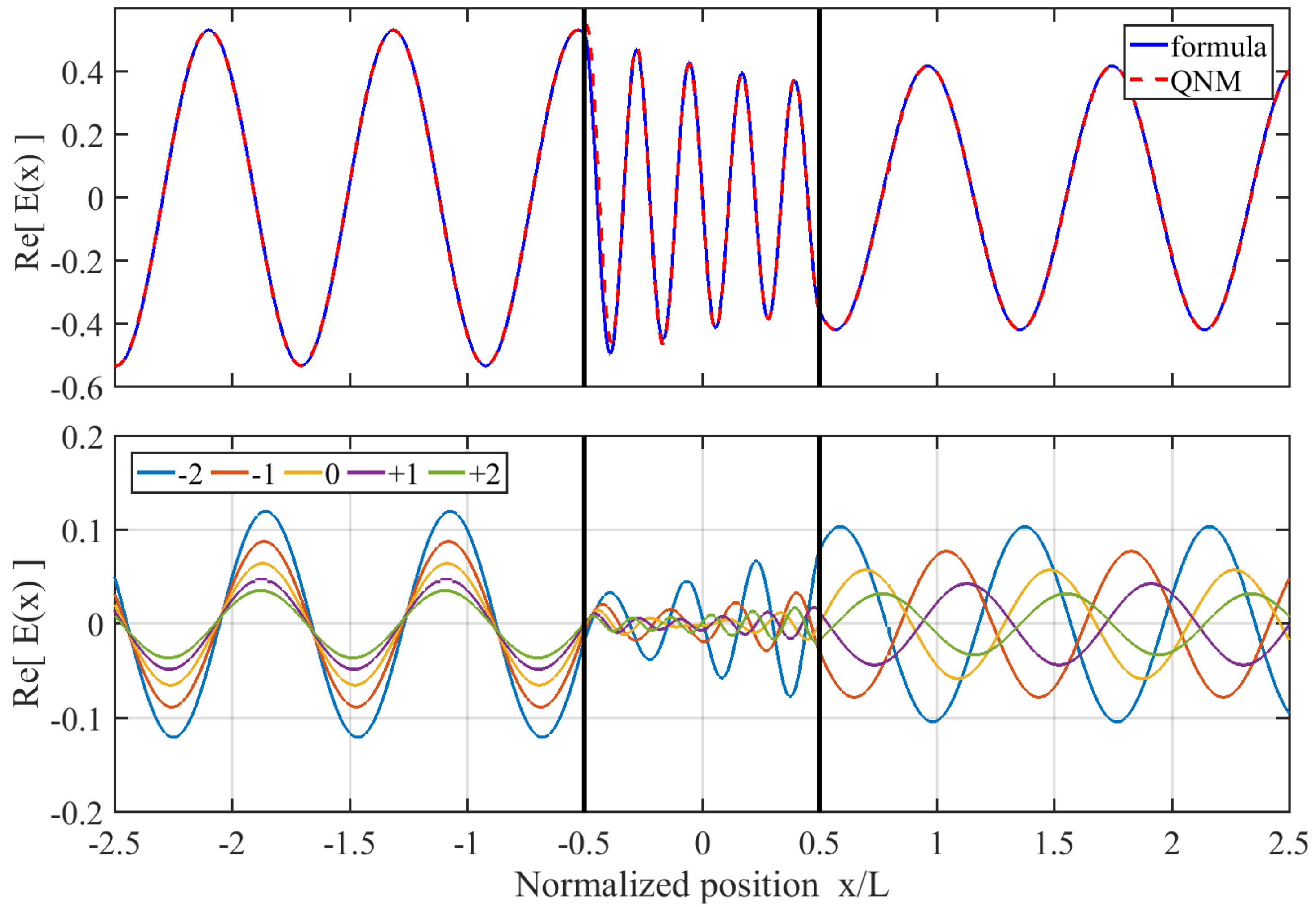}
\caption{{\cred{Top)}} The field distribution of a Lorentz-dispersive  resonator of parameters $\omega_0 L/c=10$, {\colorb{$\gamma L/c = 0.2$}}, $\Omega L/c=20$, and for a near-resonance harmonic excitation at frequency $\omega_s L / c=8$. The result of the QNMs expansion using 31 modes perfectly matches the exact formula. 
{\cred{Bottom)}} The field distribution of the five nearest  % {\colorbb{\st{normalized} }} 
QNMs to the excitation frequency.  } \label{Figws8}
\end{figure}

The field distribution, outside as well as inside the resonator, of the given test case 
%{\abs{with the absorption $\gamma L/c = 1$,}}
is presented in Fig. \ref{Figws8}  for a near-resonance excitation at $\omega_s L / c=8$. The results 
are shown for both the exact formulation  (\ref{Gx}) and the QNMs expansion  
(\ref{eq:ETexp},\ref{eq:Rexp}). The field inside the cavity is determined by {\cred{ the field continuity at the two interfaces of the resonator, i.e.,  the residues $A^q(z_q)$ 
and $B^q(z_q)$ are derived from $T^q(z_q)$ and $R^q(z_q)$ by continuity, provided that each ``conventional'' QNM \textendash \,
before the regularization factor $e^{i(z-z_q)|x+L/2|/c}$ is introduced 
by Eq. (\ref{G-QNM}) \textendash \, is a mode of the Helmholtz equation. }} 
 The {\colorb{results of the QNMs expansion perfectly}} match the exact {\cb{formulation}}{\colorb{, with a less than $2\%$ error using $31$ QNMs,}} and show no divergence {\colorb{outside the resonator and therefore}}  validate the {\cred{proposed}} approach. {\colorb{ Figure \ref{Figws8} also shows the {\colorbb{\sout{normalized} }} field distribution of the  five nearest QNMs  to the excitation frequency where the mode $0$ is the central mode (the {\colorbb{\sout{most} }} dominant mode) with the closest real frequency part to $\omega_s$.}}
% what we are saying is that the modes of Helmholtz equations (conventional QNMs) need an additional causality factor to accurately describe the response of causal open systems,
% why? not sure but may be because we derive the modes from the time (causal) Green's function, and not from the harmonic Green's function

{\cred{Figure \ref{Fig:QNMs} explicitly identifies the crucial result of this manuscript. The conventional QNMs formulation is compared to the QNMs derived taken into account the causality (\ref{G-QNM}). The field distribution of two QNMs from the previous example  is plotted up to a large distance from the resonator. The QNMs derived using causal Green's function show no divergence behavior, as predicted by the calculations, on the contrary to the ``conventional'' QNMs without the causality-related regularization factor. That finally proves that taking  into account  the causality principle  enables the construction of  well-behaved QNMs that can accurately describe the behavior of open systems.  }}

%
%%------------------------------------------------------------------------------------------------
% % Fig 5
%%------------------------------------------------------------------------------------------------
%\begin{figure}[h]
%\centering
%\includegraphics[ width=1\linewidth, keepaspectratio]{Exws2.pdf}
%\caption{ The field distribution of a Lorentz-dispersive 1-D resonator of $\omega_0=10$,  $\Omega=20$, and $L=1$ (centered at origin), for a harmonic excitation of frequency $\omega_s=2$ (nondispersive region). } \label{Figws2}
%\end{figure}

%------------------------------------------------------------------------------------------------
 % Fig 6
%------------------------------------------------------------------------------------------------

%-----------------------

\begin{figure}[h!]
\centering
\includegraphics[ width=0.9\linewidth, keepaspectratio]{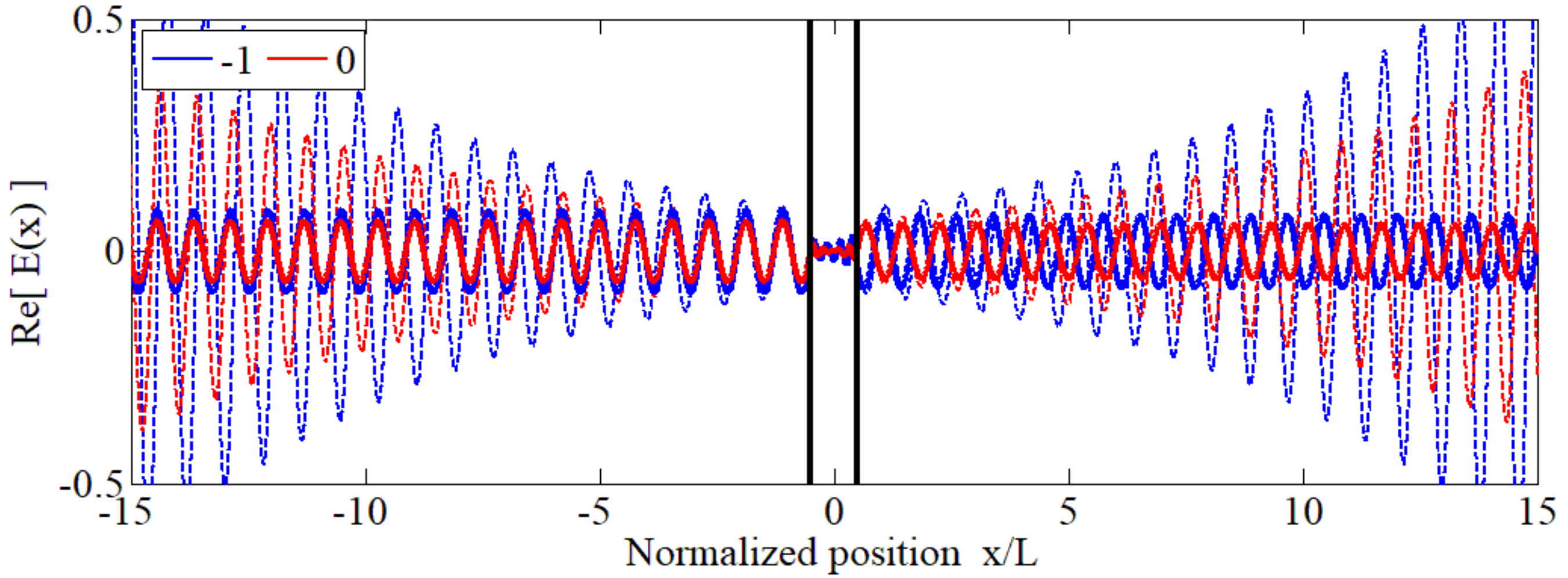}
\caption{  {\cred{ A comparison between the field distribution of QNMs formulation with the additional causality-related factor (solid lines) that shows no divergence outside the resonator, and the conventional QNMs formulation (dashed lines) that exhibits a divergence behavior. The results are shown for the two nearest QNMs to the excitation frequency of the example  in Fig. 3. }}} \label{Fig:QNMs}
\end{figure}
%-------------

%-----------------------

\begin{figure}[h!]
\centering
\includegraphics[ width=0.9\linewidth, keepaspectratio]{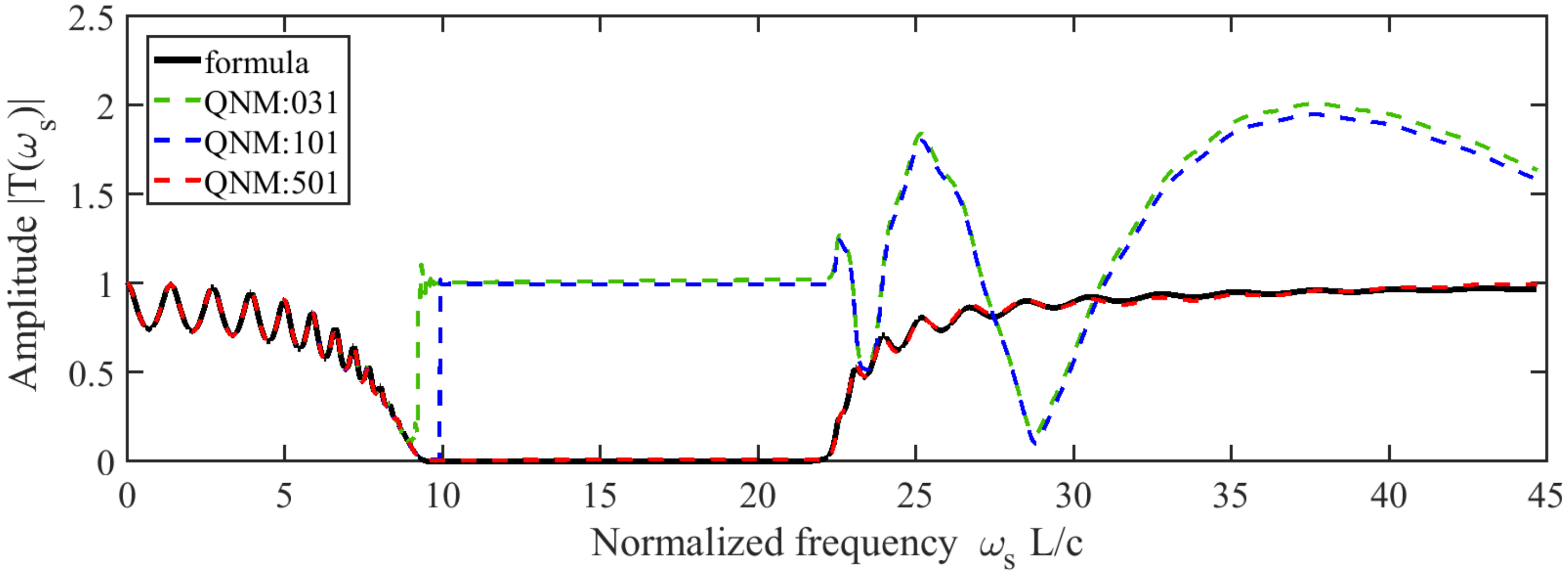}
\caption{ The exact formula for $T(z)$  and its corresponding QNMs expansion  for different excitation regions. The results are shown for different numbers of summation terms in  dashed lines. } \label{Fig:Twsabovewo}
\end{figure}
%-------------

It is worthy to note that {\colorb{the convergence is even slower for the excitation {\cb{around the resonance frequency $\om_0$ and above, as it 
can be observed for the transmission coefficient on Fig. \ref{Fig:Twsabovewo}.  That is due 
to the infinite number of modes around the resonance $\om_0$ \cite{abdelrahman2018modal} and to  the divergence of the residues $T^q$ and $R^q$ }} as $\ep(z) {\rightarrow} 1$. {\colorb{This divergence is then compensated by the $1/z_q$ {\cred{factor}} {\cb{in (\ref{eq:ETexpbefore}), but}} }} in this case it is needed to include many summation terms to ensure convergence}}. {\cb{\sout{An alternative solution to improve the 
convergence is to introduce a small $\delta\ep$  to the permittivity $\ep(z)$ while evaluating the residues and that should ensure the convergence. }}}

\section{Conclusion}
The  quasi-normal modes (QNMs) expansion of the Green's function is derived using complex analysis {\cred{and}} taking into consideration the frequency dispersion, the causality 
principle and thus the finite velocity of {\cred{electromagnetic}} waves. The resulting formulation of the 
QNMs expansion shows no divergence of the fields when $|\r| \to \infty$, {\colorb{{\cb{leading to}} the construction of {\cred{well-behaved}} modes of the open system that are able to accurately describe the fields distribution inside and outside the open resonator.}} {\cred{In this approach, the regularization process  (the additional  factor) is intrinsic and appears when the QNMs are redefined using the causal Green's function formulation.}} A simple one-dimensional homogeneous {\cb{absorptive}} medium with a Lorentz 
frequency dispersion is considered to validate the general derivation and the arguments on 
which it is based. The fields evaluated using the QNMs expansion, inside and outside 
the resonator,
perfectly match the exact expressions {\colorb{ if enough summation terms are considered}}. %{\abs{, even with absorption}}. 
These results bring new elements to show that 
the QNMs can form a complete set to express in the whole space the electromagnetic fields 
diffracted by dispersive {\cb{and absorptive}} 
materials. {\cb{The general derivation is 
applicable to {\cred{systems with a discrete set of resonances 
like bounded}} scatterers in two and three dimensions.}} 

\section*{Funding}
This work was supported by the French National Agency for Research (ANR) under 
the project ``Resonance'' (ANR-16-CE24-0013). We express our gratitude to all the participants in the project ``Resonance'' and to Prof. Aladin Hassan Kamel (Ain-Shams University, Egypt) for the valuable discussions.

%%%%%%%%%%%%%%%%%%%%%%% References %%%%%%%%%%%%%%%%%%%%%%%%%

%%%%%%%%%% If using BibTeX:

%%%%%%%%%% If preparing manually:
% \newcommand{\enquote}[1]{``#1''}

% \bibitem{Zhang:14}
% Y.~Zhang, S.~Qiao, L.~Sun, Q.~W. Shi, W.~Huang, L.~Li, and Z.~Yang,
%   \enquote{Photoinduced active terahertz metamaterials with nanostructured
%   vanadium dioxide film deposited by sol-gel method,}
%   {\protect\JournalTitle{Optics Express}} \textbf{22}, 11070--11078 (2014).

% {Optical Society}, \enquote{{OSA Publishing},}
%   \url{http://www.osapublishing.org}.

% \bibitem{FORSTER2007}
% P.~Forster, V.~Ramaswamy, P.~Artaxo, T.~Bernsten, R.~Betts, D.~Fahey,
%   J.~Haywood, J.~Lean, D.~Lowe, G.~Myhre, J.~Nganga, R.~Prinn, G.~Raga,
%   M.~Schulz, and R.~V. Dorland, \enquote{Changes in atmospheric consituents and
%   in radiative forcing,} in \enquote{Climate Change 2007: The Physical Science
%   Basis. Contribution of Working Group 1 to the Fourth assesment report of
%   Intergovernmental Panel on Climate Change,}  S.~Solomon, D.~Qin, M.~Manning,
%   Z.~Chen, M.~Marquis, K.~B. Averyt, M.~Tignor, and H.~L. Miler, eds.
%   (Cambridge University Press, 2007).

% \end{thebibliography}

\end{document}